**Title: The Importance of Justified Patient Trust in unlocking AI's potential in mental healthcare**

Tita Alissa Bach[1]*; Männikkö, Niko[2,3]*

Affiliations:

[1]Group Research and Development, DNV, Høvik, Norway

[2]Centre for Research and Innovation, Oulu University of Applied Sciences, Oulu, Finland

[3]Research Unit of Health Science and Technology, Faculty of Medicine, University of Oulu, Finland

*Both authors contributed equally to this work

Correspondence: tita.alissa.bach@dnv.com

**Introduction**

*"Mental health is a basic human right." WHO mental health (1)*

Mental health, as defined by the World Health Organization (WHO), is "a state of mental well-being that enables people to cope with the stresses of life, realize their abilities, learn well and work well, and contribute to their community" (1). In recent years, mental health challenges have surged globally, with rising cases and increasing complexity. Simultaneously, the shortage of mental healthcare providers has become more acute, creating gaps in care (2). Artificial Intelligence (AI)-enabled systems[1] or, AI systems in short, have the potential to revolutionize mental healthcare by addressing these gaps, offering solutions that range from digital diagnostics to therapeutic tools[1] (3). Specifically, AI-enabled systems have been used to help mental healthcare by directly interacting with patients on, for example, self-management mobile health apps in treating depression, anxiety, post-traumatic stress disorders, sleep disorders, and suicidal behaviors (4,5), as well as diagnosing users with behavior or responses associated with mental health conditions, developing risk profiles, and deploying context-specific interventions (6).

However, the success of these AI-driven innovations hinges on one crucial factor: patient trust. Without trust, patients may hesitate to engage with AI systems, limiting the technology's impact. Real-world cases have already highlighted the risks of undermining this trust. For instance, the National Eating Disorders Association (NEDA) recently removed AI chatbot, Tessa, from a support hotline after concerns arose that it was providing harmful advice, potentially exacerbating the conditions of vulnerable users who were patients of eating disorders (7). Similarly, Sonde Health's voice analysis AI, which uses vocal biomarkers to assess depression, has been criticized for overlooking the diverse speech patterns of non-typical users, such as those with disabilities or regional and non-native speech differences (8). In addition, patient concerns about data privacy and potential biases in AI systems, how patient data is used, and the potential for AI systems to perpetuate existing inequalities have been reported as key trust barriers (2). These examples highlight the fragility of trust in AI systems, particularly in the sensitive domain of mental health, where patient vulnerability is already high at baseline (9).

Trust is delineated as the "willingness to render oneself vulnerable" to a capability, founded on an evaluation of congruence in intentions or values (10). Trust relationships can be established among individuals and between individuals and technology (11). Trust is often described as a connection between a trustor and a trustee, with the hopeful anticipation that the trustee will meet the trustor's expectations (12). Trust relationships usually do not have legally binding obligations and are therefore susceptible to deceit. As a result, various factors contribute to and affect the dynamic of trust relationships.

---

[1]Any system that contains or relies on one or more AI components. AI components are distinct units of software that perform specific functions or tasks within an AI-enabled system. They consist of a set of AI models, data, and algorithms, which, through implementation, create the AI component (27)

In this paper, we focus specifically on trust in AI-enabled systems by mental health patients who are also the direct users of these systems, highlighting the most sensitive and direct relationship between AI systems and those whose mental healthcare is impacted by them. In mental healthcare, AI systems can be used by mental health professionals (13,14), patients, and patients' families or caregivers (15). However, when patients are not the direct users of these systems, their trust in them is likely to be indirect and mediated by their trust in the healthcare professionals or family members who utilize AI systems on their behalf. Patients as users have different trust needs and significantly higher risks in using these systems than do non-patient users. Trust in this context is thus delicate, as patients' emotional and cognitive states may make them particularly vulnerable to the risks of over-reliance on AI-enabled systems. Most importantly, patients are usually the primary stakeholders of many AI-enabled mental health applications, irrespective of who the users may be (16). The user experience for patients is deeply personal, and their trust in these systems directly influences their engagement and, ultimately, the outcomes of their care. This is because patient commitment to actively engage in the care plan is the single most critical determinant of positive outcomes (6). Therefore, building patient trust is not just beneficial for empowering patients and giving them a sense of control over their care, it is absolutely vital for ensuring successful and meaningful care outcomes (6).

In some cases, giving patients access to use AI systems without any support from mental healthcare professionals or caregivers requires careful consideration (17). Such cases arise when patients are deemed clinically incapable of making their own decisions. This complicates the trust equation, as patients may no longer be seen as users of an AI system even if they interact directly with it. As a result, their caregivers may be viewed as the users.

**Fostering justified trust**

Previous studies have shown factors influencing user trust in AI systems that are related to individual characteristics, such as user personality and cultural backgrounds (18,19). These factors are likely to be more difficult to change; thus, they should be taken into consideration while developing and deploying an AI system (10). In contrast, there are factors that are easier to improve in order to foster user trust in AI systems. For example, preparing and adjusting the socio-ethical environments where the AI systems are (to be) deployed, improving user acceptance to AI systems, managing user needs and expectations, and improving user perceptions (10). These factors can be improved more easily by, for example, providing users with more transparency in how an AI system is developed and how the AI system works (2).

While patients' lack of trust may slow the adoption of AI systems in mental healthcare, a more significant concern arises when patients "trust incorrectly"(20). Initial hesitation or scepticism is a natural and expected reaction when humans encounter new or unfamiliar technology, making it easier to anticipate and address. However, the risks associated with overtrust or blind trust in AI systems are less frequently discussed (21), which could lead to serious consequences within mental healthcare.

Trusting correctly means that the trust is justified and based on evidence, knowledge, experiences and/or skills (20,22,23). In mental healthcare, it means that although patients trust an AI system, the patients need to keep their critical thinking while interpreting the output of the AI system. Such critical thinking can be encouraged by providing patients with skills to understand AI system's capabilities and limitations (2), so that patients can recognize deviations of AI systems' operations and output. Similarly, providing patients with tools to control their data can encourage patients' critical thinking (2). In care treatments where the relationships between patient-professionals are fundamental to positive outcomes, such as in psychotherapy (24,25), it is important to also provide patients with access to mental health professionals, alongside patients' use of AI systems (25).

However, at the moment, there is still too little research investigating the effectiveness of various AI systems in mental healthcare to build evidence (6). This is unfortunate and does not help patients form justified trust through collected scientific evidence. Forming justified trust thus needs to depend on users' and domain experts' experiences, knowledge and/or skills with the hope that over time they can build evidence on the positive and negative effects of an AI system on patients' mental health (20,22,23).

**Maintaining justified trust over time**

User trust in AI systems is dynamic and can change over time (10,26). A review by Cabiddu et al. (2022) suggest that *initial user trust* development in AI systems is influenced by users' propensity to trust, human-like features, and the user-perceived usefulness of AI systems (26). It is suggested that users often attribute human traits to AI systems, as human-like features enhance emotional connections and relatability, fostering trust. For example, AI systems that mimic human communication styles can create a sense of familiarity, making users more comfortable interacting with them and ultimately trusting their output.

In addition to users' initial trust in AI systems, the review suggests that there are other factors as well that influence *user trust over time*, including social influence, familiarity, and system-like features (26). Regarding the system-like features, user trust is suggested to be significantly influenced by the users' perceived reliability and predictability of the AI system to perform its intended task. This suggests that users base their trust on evaluating whether their initial perception of the usefulness of the AI system aligns with its actual performance after repeated use over time. Such evaluation is in line with fostering justified trust as the trust is rooted in the user knowledge and experiences built and collected over time (22).

As users become more familiar with the AI systems, especially if they have strong social support to continue the use, as well as a positive perception of its usefulness through continued reliable and predictable AI output, sustained user trust is established. In such a scenario, established user trust can nevertheless still change into user distrust or mistrust, for example, when AI systems make errors that impact users directly, or overtrust, for example, when users under time pressure and/or with low cognitive capacity act upon AI output without any evaluation or judgment (27).

To maintain justified trust, it is crucial to continually promote critical thinking so that users may base their evaluation rooted in collected evidence, if any, as well as knowledge, experiences, and/or skills. Accordingly, patient education on AI's capabilities and limitations is extremely valuable for maintaining justified trust, as well as incorporating feedback from patients to improve the AI system in question. The downside of maintaining justified trust in this manner is that it requires a high cognitive load, and it depends on the patients' ability to think critically each time. This can become an issue for mental health patient as patients may use AI systems in their most vulnerable conditions, when their cognitive capacity is likely limited.

Therefore, it is only ethical and responsible to develop, deploy, and continuously improve AI systems together with patients (27), especially to understand what influences patients' cognitive capacity and critical thinking when using AI systems. It is crucial to match specific user populations' characteristics and needs to the design of AI systems, specifically AI interface and features where human-AI interaction happens (10,27). For example, this can be done by identifying users' needs to determine which key aspects of AI output are to be displayed, or not, in the interface.

An AI system to be used by patients with sensory sensitivity, for example, should be designed with fit-for-purpose visual and audio elements by avoiding bright colours, loud noises, and an overstimulating display. AI systems developed for patients with PTSD or trauma can, for example, gradually introduce more and more challenging topics as trust is built and patients are more comfortable, rather than immediately overwhelming patients with sensitive materials. AI systems can also incorporate customizable trigger detection such that patients can specify any words, topics, or stimuli that they find distressing to let AI systems adjust themselves. All these examples show the importance of embedding an

AI feature to personalize user interface based on users' preferences, as well as giving control to users to decide what, how much, how, and when preferred information is to be presented to them (27). Importantly, such personalization can help patients to evaluate AI output without requiring additional workload and within their cognitive capacity at the time of use, maintaining their justified trust. One size thus does not fit all.

**Conclusions**

Given that mental healthcare already presents unique ethical and legal challenges, the additional use of AI systems requires scrutiny and appropriate, fit-for-purpose regulation (17). The role of regulators here is key to aligning practices of development through the use of AI systems with responsible and ethical principles (17). For example, ensuring that the claimed benefits of using AI systems are true and not misleading lay users, specifically those made by for-profit AI vendors and suppliers. In addition, because the use of AI systems in mental healthcare is still a relatively new phenomenon, stakeholders in the field need to be provided with channels and platforms to interact and share learning, helping to identify the obstacles and enablers of justified trust-fostering based on the specific benefits and risks of AI systems to patients (28). In conclusion, detailed personalization based on the patients' mental health conditions is fundamental to ensuring that AI systems provide effective care treatment as intended while maintaining user justified trust.

**Data availability statement**

The original contributions presented in the study are included in the article, and further inquiries can be directed to the corresponding author.

Author contributions

TAB: Writing – original draft, Writing – review & editing.

NM: Writing – original draft, Writing – review & editing.

**Funding**

The author(s) declare that no financial support was received for this article's research, authorship, and/or publication.

**Conflict of interest**

The author declares that the research was conducted without any commercial or financial relationships that could potentially create a conflict of interest.